\documentstyle[12pt]{article}

\newcommand{\be}{\begin{equation}}
\newcommand{\ee}{\end{equation}}
\newcommand{\bea}{\begin{eqnarray}}
\newcommand{\eea}{\end{eqnarray}}

\def\pa{\partial}                      %
\def\tr{{\mathrm Tr\,}}                %
\def\G{{\cal G}}                       %
\def\Z{{\bf Z}}                        %
\def\M{{\cal M}}                       %
\def\H{{\cal H}}                       %
\def\A{{\cal A}}                       %
\def\F{{\cal F}}                       %
\def\D{{\cal D}}                       %
\def\R{{\bf R}}                        %
\def\cR{{\cal R}}                      %
\def\wt{\widetilde}                    %
\def\ad{{\mathrm ad }}                 %

\begin{document}

\thispagestyle{empty}
\setcounter{page}{0}

\renewcommand{\thefootnote}{\fnsymbol{footnote}}
\fnsymbol{footnote}

\vspace{.4in}

\begin{center} \Large\bf 
Classical Wakimoto Realizations of Chiral WZNW Bloch Waves

\end{center}

\vspace{.1in}

\begin{center}
J.~Balog$^{(a)}$, L.~Feh\'er$^{(b)}$\protect\footnote{
Corresponding author's e-mail: lfeher@sol.cc.u-szeged.hu, 
phone/fax: (+36) 62 544 368.}
and L.~Palla$^{(c)}$ \\

\vspace{0.2in}

$^{(a)}${\em  Research Institute for Nuclear and Particle Physics,} \\
       {\em Hungarian Academy of Sciences,} \\
       {\em  H-1525 Budapest 114, P.O.B. 49, Hungary}\\

\vspace{0.2in}       

$^{(b)}${\em Institute for Theoretical Physics,
        J\'ozsef Attila University,} \\
       {\em  H-6726 Szeged, Tisza Lajos krt 84-86, Hungary }\\

\vspace{0.2in}   

$^{(c)}${\em  Institute for Theoretical Physics, 
Roland E\"otv\"os University,} \\
{\em  H-1117, Budapest, P\'azm\'any P. s\'et\'any 1 A-\'ep,  Hungary }\\

\end{center}

\vspace{.2in}

\begin{center} \bf Abstract
\end{center}

{\parindent=25pt
\narrower\smallskip\noindent
It is well-known that the chiral WZNW Bloch waves satisfy a 
quadratic classical exchange algebra which implies the affine Kac-Moody
algebra for the corresponding currents.
We here obtain a direct derivation of the exchange algebra
by inverting the symplectic form on the space of Bloch waves,
and give a completely algorithmic construction 
of its generalized free field realizations that extend  
the classical Wakimoto realizations of the current algebra.  
}

\vspace{11 mm}
\begin{center}
keywords: WZNW model, exchange algebra, Wakimoto realizations
\end{center}
\vfill\eject

\section{Introduction}
\setcounter{equation}{0}

\renewcommand{\thefootnote}{\arabic{footnote}}
\setcounter{footnote}{0}

The Wess-Zumino-Novikov-Witten (WZNW) model \cite{Witt}
of conformal field theory  has proved to be the source 
of interesting structures that play an increasingly important r\^ole
in theoretical physics and in mathematics \cite{DiF,EFV}.
Among these structures are the quadratic Poisson bracket algebras 
\cite{Bab}--\cite{BFP} 
that arise from the chiral separation of the degrees of freedom 
in the model, which is based on the form of the classical solution 
of the field equation given by
\be
g(x_L, x_R) = g_L(x_L) g_R^{-1}(x_R).
\ee
Here $x_C$ $(C=L,R)$ are lightcone coordinates and the $g_C$ 
are quasiperiodic group valued fields with equal monodromies,
$g_C(x+2\pi)=g_C(x) M$ for some $M$ in the WZNW group $G$. 
The classical exchange algebras can be regarded as fundamental  
since the affine Kac-Moody (KM) symmetry follows as their consequence, and 
help to better understand the quantum group properties 
of the model 
(see e.g.~\cite{qgroup}) by means of canonical quantization 
\cite{Chu2,FHT,CL}.
One should however note that after the chiral separation,
which essentially amounts to forgetting the equal monodromy constraint
on the pair $(g_L, g_R)$,  
the Poisson structure is highly 
non-unique \cite{Gaw,FG,BFP}.
A natural choice consists in restricting  the monodromy $M$ of the 
chiral WZNW fields to be
diagonal, in which case the resulting classical exchange algebra  can be
described as follows \cite{BDF,Chu5,BBT}.
Denoting any of the chiral WZNW `Bloch waves' by $b(x)$,
\be
b(x+2\pi) = b(x) e^{\omega},
\qquad
\omega \in {\cal H},
\ee
where ${\cal H}$ is a Cartan subalgebra of the Lie algebra $\G$ of $G$,
the exchange algebra reads as 
\be 
\Big\{b(x)\stackrel{\otimes}{,} b(y)\Big\}
={1\over\kappa}\Big(b(x)\otimes
b(y)\Big)\Big({\cal R}(\omega) 
+{1\over 2}{\hat I} \,{\rm sign}\,(y-x)\Big), 
\quad 0< x, y<2\pi,
\label{xcha}\ee 
\be
{\cal R}(\omega)= {1\over 2}\sum_{\alpha\in \Phi} 
\coth({1\over 2}\alpha(\omega)) 
E_\alpha\otimes E^\alpha,
\qquad 
\hat I= E_\alpha \otimes E^\alpha + H_k \otimes H^k,
\label{Romega}\ee
where $E_\alpha, H_k$ is a Cartan-Weyl basis of $\G$ 
and $\kappa$ is a constant. 
For the `exchange r-matrix' ${\cal R}(\omega)$ to be non-singular,  
the monodromy parameter $\omega$ has to be  restricted to a 
domain in ${\cal H}$ where $\alpha(\omega)\notin i 2\pi  {\bf Z}$
for any root $\alpha$. 
Notice also from (\ref{xcha}) that the $2\pi$-periodic current 
\be
J= \kappa b' b^{-1}
\label{Jdef}\ee
indeed satisfies the KM Poisson brackets
\be
\{ \tr(T_aJ)(x), \tr(T_b J)(y)\}= \tr([T_a, T_b] J)(x)\delta(x-y) 
+ \kappa\tr(T_aT_b) \delta'(x-y),
\label{JPB}\ee
where $\delta(x-y)= {1\over 2\pi} \sum_{n\in \Z} e^{in (x-y)}$ 
and $T_a$ is any basis of $\G$ 
(for the notations, see also sec.~2). 

Our main purpose here is to construct generalized 
free field realizations 
of the exchange algebra given by (\ref{xcha}) 
that extend the Wakimoto realizations \cite{Wa}
of the current algebra. 
The Wakimoto realizations of the KM current 
are well-understood both at the classical and quantized level 
and proved useful in many 
respects (see e.g.~\cite{DiF,FF,BMP,GMOMS,dBF}).   
Their extension to accompanying realizations of the WZNW Bloch waves 
have also received attention \cite{AS,IY,FG,GMOMS}, 
but, except for the simplest cases, 
a fully algorithmic construction has not appeared yet.
At the classical level, 
we present such a construction  in sec.~3.
The construction is summarized by the diagram (\ref{diagram}), 
where all the arrows represent Poisson maps,
which is a new result for the map $\widehat W$ defined by (\ref{wakiext}).
For completeness, in sec.~2 we also give  
a derivation of the exchange algebra.
This  derivation is  
quite different from those in \cite{BDF,BBT} and conceptually close 
to that in \cite{Chu5}.
In order to obtain (\ref{xcha}), we will directly 
invert the symplectic form -- which we use in sec.~3 as well --  
that the space of Bloch waves inherits from 
the full WZNW phase space.
 
\section{Derivation of the exchange algebra of Bloch waves}
\setcounter{equation}{0}

Throughout the paper, let $\G$ be either a complex 
simple Lie algebra or its normal real form, 
and $G$ a corresponding Lie group. 
Denote by  $\widetilde G$  (resp.~$\widetilde \G$) the loop group 
(algebra) consisting 
of the $2\pi$-periodic, smooth, $G$-valued ($\G$-valued) 
functions on the real line $\R$.
Choose a Cartan subalgebra $\H\subset \G$ that admits the   
root space decomposition
\be
\G=\H \oplus \sum_{\alpha\in \Phi} \G_\alpha,
\ee
and an associated basis $H_k\in \H$, $E_\alpha \in \G_\alpha$ 
normalized by
$\tr\left(E_\alpha E_{-\alpha}\right) = {2\over \vert \alpha\vert^2}$,
where $\tr$ denotes an invariant scalar product on $\G$.
By using this basis any $A\in \G$ can be decomposed as  
\be
A=A^0+ A^r
\quad\hbox{with}\quad 
A^0\in \H,\quad 
A^r=\sum_{\alpha\in \Phi} E_\alpha  \tr(E^\alpha A),
\quad
E^{\alpha}:= {1\over 2}{\vert \alpha \vert ^2 } E_{-\alpha} .
\label{0rdecomp}\ee
Fix an open domain $\A\subset \H$ which has the properties that 
$\alpha(\omega)\notin i2 \pi  {\bf Z}$ for any root, 
$\alpha\in \Phi\subset {\cal H}^*$,
and the map $\A\ni \omega\mapsto e^{\omega} \in G$ is injective.

Let us now define $\M_{Bloch}^G$ by
\be
\M^{G}_{Bloch}:=\{ b \in C^\infty(\R, G)\,\vert\,
b(x+2\pi)= b(x) e^{ \omega},
\quad 
\omega\in \A \subset \H\,\},
\label{Bloch}\ee
and endow it with the differential 2-form 
\be
\Omega_{Bloch}^{G,\kappa}(b)=
- {\kappa\over 2 }\int_0^{2\pi}dx\, 
 {\tr}\left(b^{-1}d b \right) \wedge
\left(b^{-1}db \right)'
-{\kappa\over 2 }  {\tr}\left((b^{-1} d b)(0)\wedge d \omega\right).
\label{Blochform}\ee 
It is easy to check that $d\Omega_{Bloch}^{G,\kappa}=0$ and 
below we shall see that this formula actually defines a symplectic 
form on $\M_{Bloch}^G$.
Note that $\Omega_{Bloch}^{G,\kappa}$ can be obtained from the natural 
symplectic form 
on the space of classical solutions of the WZNW model, 
as explained in \cite{Chu5,Gaw,FG}.

It will be convenient to parametrize $b\in \M_{Bloch}^G$ as 
\be
b(x) = h(x) \exp\left(x \bar\omega \right),
\qquad \bar \omega:= {\omega\over 2\pi},
\label{Blochpar}\ee
where $\omega\in \A$ and $h\in \widetilde{G}$.
This one-to-one parametrization yields the identification 
\be
\M_{Bloch}^G =\widetilde{G}\times \A=\{ (h, \omega)\}.
\ee
Correspondingly, 
a vector field $X$ on $\M^G_{Bloch}$ is parametrized by
\be
X=(X_h, X_\omega)
\qquad X_h \in T_h \wt{G}
\qquad
X_\omega \in T_\omega \A \simeq \H
\ee
with $h^{-1} X_h \in T_e \wt{G} \simeq \wt{\G}$.
By regarding $\omega$ and $h$ as evaluation functions 
on $\M^G_{Bloch}$, we may write 
$X_\omega=X(\omega)$ and $X_h(x) = X(h(x))$.
Equivalently, $X$ can be characterized by its action on $b(x)$,
\be
b^{-1}(x) X(b(x))=e^{-x\bar \omega } h^{-1}(x) X(h(x)) e^{x\bar \omega} 
+ x X(\bar\omega),
\label{Xb}\ee
where the function $b^{-1}(x) X(b(x))$ on ${\bf R}$ is
uniquely determined by its restriction to $[0, 2\pi]$.
In general, the derivative $X(F)$ of a function $F$ on $\M^G_{Bloch}$ 
is defined by using that any vector is the velocity to a smooth curve. 
That is, if the value of the vector field $X$ at $b\in \M^G_{Bloch}$ 
coincides with the velocity to the curve  $\gamma (x,t)$ at $t=0$, 
$\gamma (x,0)=b(x)$, then for a 
differentiable  function $F$ we have 
$X(F)[b]=\frac{\rm d}{{\rm d}t}F[\gamma (x,t)]\vert_{t=0}$.  

A function $F$ on $\M^G_{Bloch}$ is henceforth called {\em admissible} 
if its derivative with respect to any vector field $X$ 
exists and has the form 
\be
X(F) = \langle dF,X \rangle=\tr( d_\omega F X_\omega)+
\int_0^{2\pi} dx\, \tr\left( (h^{-1} d_h F) (h^{-1} X_h)\right)
\ee
where 
\be
dF=(d_h F, d_\omega F)
\quad
\hbox{with}\quad 
d_h F\in T_h^* \widetilde{G},
\quad
d_\omega F\in T^*_\omega \A
\ee
is the exterior derivative of $F$.
In this definition we identify $T^*_\omega \A$ with $\H$ by means of the scalar
product ``$\tr$'' and also identify $T_e^* \widetilde G$ 
with $\widetilde{\G}$ by the scalar product $\int_0^{2\pi} \tr(\cdot,\cdot)$,
whereby we have $h^{-1} d_h F\in T_e^*\widetilde{G}=\widetilde{\G}$.
It is clear that the local evaluation functions $\M^G_{Bloch}\ni b\mapsto b(x)$
are differentiable but not admissible, while e.g.~the Fourier components of the current 
$J$ and the components of $\omega$ are admissible functions.
It is also worth noting that the matrix elements of the nonlocal 
$G$-valued function
given by the path ordered exponential integral of the current $J$ over 
a period are,
in fact, not admissible, but the trace of any of the powers of this 
(Wilson loop) function
is admissible, since it can be expressed as a function of $\omega$ alone.  

We now wish to show that
$\Omega_{Bloch}^{G,\kappa}$ is symplectic 
in the sense that it permits 
to associate a unique hamiltonian vector field, $Y^F$,  
with any  admissible function, $F$.
The defining property of $Y^F$ is that it must satisfy 
\be
\langle dF, X\rangle =X(F)=\Omega^{G,\kappa}_{Bloch}(X,Y^F)
\label{YFdef}\ee
for any vector field $X$.
In order to determine $Y^F$ from this equality,
we  first point out that in terms of the variables $(h,\omega)$
\bea
\Omega_{Bloch}^{G,\kappa} (h,\omega)&=&
-{\kappa\over 2} \int_0^{2\pi}dx\, \tr\Bigl(
(h^{-1} dh)\wedge (h^{-1} dh)' \nonumber\\
 &+&  2\bar \omega (h^{-1} dh)\wedge (h^{-1} dh)
-2d\bar \omega \wedge h^{-1} dh \Bigr).
\eea
It follows that 
\begin{eqnarray}
\Omega^{G,\kappa}_{Bloch}(X,Y^F)&= & -\kappa\int_0^{2\pi}
\tr\Bigl( 
(h^{-1} Y^F(h))'h^{-1} X(h)  + \bar \omega [h^{-1} X(h), h^{-1} Y^F(h)]
\nonumber\\
& + & Y^F(\bar \omega) h^{-1} X(h) -X(\bar\omega) h^{-1} Y^F(h) \Bigr).
\end{eqnarray}
This implies the  following equations for $Y^F$: 
\bea
\left(h^{-1} Y^F(h)\right)' + [h^{-1} Y^F(h), \bar\omega] 
+ Y^F(\bar\omega)&=&
-{1\over \kappa} h^{-1} d_h F
\label{G1}\\
\int_0^{2\pi}dx\, (h^{-1} Y^F(h))^0 &=& {2\pi\over \kappa}  d_\omega F.
\label{G2}\eea
Given $d_h F$ and $d_\omega F$, we will next find $b^{-1} Y^F(b)$.

On account of (\ref{Xb}), 
(\ref{G1}) is in fact equivalent to 
\be
\left( b^{-1} Y^F(b)\right)'(x)= -{1\over  \kappa} 
e^{-\bar\omega x} (h^{-1} d_h F)(x)
e^{\bar\omega x},
\label{G4}\ee
whose solution   is given by
\be
b^{-1}(x) Y^F(b(x))= b^{-1}(0)Y^F(b(0)) 
- {1\over \kappa} \int_0^x dy\, e^{-\bar\omega y} (h^{-1} d_h F)(y) 
e^{\bar\omega y}.
\label{G5}\ee
Hence the only  nontrivial problem  is to determine the initial value
\be
Q_F:= b^{-1}(0) Y^F(b(0)) = h^{-1}(0) Y^F(h(0)).
\label{G6}\ee
To this end, note from (\ref{Xb}) that
\be
Y^F(\omega)=
e^{\omega} b^{-1}(2\pi) Y^F(b(2\pi)) e^{- \omega} -b^{-1}(0) Y^F(b(0)).
\label{G8}\ee
By using (\ref{G5}), 
the Cartan part of (\ref{G8})  requires that
\be
Y^F(\omega)= -{1\over \kappa} \int_0^{2\pi} dx\, (h^{-1} d_h F)^0(x),
\label{G9}\ee
while the root part of (\ref{G8}) gives
\be
e^{-\omega} Q_F^r e^{ \omega} -Q_F^r= -{1\over \kappa} 
\int_0^{2\pi} dx\, e^{-\bar \omega x} (h^{-1} d_h F)^r(x) e^{\bar \omega x},
\label{G10}\ee
where $Q_F=Q_F^0+Q_F^r$ 
according to (\ref{0rdecomp}).
Then  (\ref{G10}) completely determines $Q_F^r$ as
\be
Q_F^r={1\over \kappa}\sum_{\alpha\in \Phi} 
{E_\alpha\over 1-e^{-\alpha(\omega)}}
\int_0^{2\pi} dx\, e^{-\alpha(\bar\omega)x} \tr
\left( (h^{-1} d_h F)(x) E^\alpha\right).
\label{G11}\ee
As for the remaining unknown, $Q_F^0$, (\ref{G2}) with (\ref{Xb}) 
and (\ref{G8}) 
leads to the result:
\be
2\pi \kappa Q_F^0=2\pi d_\omega F -\pi \int_0^{2\pi} dx\,
(h^{-1} d_h F)^0(x) +\int_0^{2\pi} dx\, \int_0^x dy\,
(h^{-1} d_h F)^0(y).
\label{G12}\ee  

In conclusion, we have found that 
the hamiltonian vector field $b^{-1} Y^F(b)$ is uniquely 
determined and is explicitly given by (\ref{G5}) with $b^{-1}(0) 
Y^F(b(0))=Q_F$ 
in (\ref{G11}), (\ref{G12}).
In the derivation of $Y^F$ we have crucially used that $\omega$ is restricted
to the domain $\A\subset \H$.
At the excluded points of $\H$ some denominators in (\ref{G11}) 
may vanish, whereby $\Omega_{Bloch}^{G,\kappa}$ becomes singular.

The Poisson bracket of two `smooth enough' admissible functions
$F_1$ and $F_2$ on $\M_{Bloch}^G$ is determined by the formula 
\be
\{ F_1,F_2\}=Y^{F_2}(F_1)=\Omega_{Bloch}^{G,\kappa}(Y^{F_2}, Y^{F_1}).
\ee
We now explain that in a certain sense this Poisson bracket is encoded 
by the classical exchange algebra  (\ref{xcha}).
For this purpose we consider functions of the form 
\be
F_{\phi}(h, \omega)= \int_0^{2\pi} dx\, \tr( \phi(x) b^\Lambda(x) ),
\ee
where $b^\Lambda(x)$ is taken in a  
representation\footnote{We use the notation 
$\tr= c_\Lambda {\rm{tr}}_\Lambda$,
where $\rm{tr}_\Lambda$ is the trace over the representation 
$\Lambda$ and $c_\Lambda$ is a normalization factor that makes 
$c_\Lambda \rm{tr}(A^\Lambda B^\Lambda)$
 independent of $\Lambda$ for $A,B\in\G$.}
$\Lambda$ of $G$ 
and $\phi(x)$ is a smooth, matrix valued,
smearing-function in that representation.
It is easy to check that $F_\phi$ is admissible if  
\be
\phi^{(k)}(0)=\phi^{(k)}(2\pi)= 0 \quad \forall k=0,1,2\ldots\,,
\label{G13}\ee
and the exterior derivative of $F_\phi$ at $(h,\omega)$ is given by  
\be
(d_\omega F_\phi)(h,\omega)={1\over 2\pi} 
\sum_k H^k \tr\left( H^\Lambda_k \int_0^{2\pi} dx\, 
( x \phi(x) b^\Lambda(x) )\right),
\label{G14}\ee
\be
\left( (h^{-1} d_h F_\phi)(h,\omega)\right)(x)=
\sum_a T^a \tr\left( 
\phi(x) h^\Lambda(x)T^\Lambda_a e^{x \bar\omega^\Lambda} \right)
\quad\hbox{for}\quad
x\in [0, 2\pi].
\label{G15}\ee
We here denote by $H_k$, $H^k$ and $T_a$, $T^a$ dual bases of $\H$ 
and $\G$, respectively.
The last formula extends to a smooth $2\pi$-periodic function 
on the real line
precisely if (\ref{G13}) is satisfied.      
The hamiltonian vector field
$Y^{F_\phi}$ is then found to be 
\be
\left( b^{-1} Y^{F_\phi}(b)\right)(x)= Q_{F_\phi} -
{1\over \kappa} \sum_a T^a \int_0^x dy\, 
\tr(\phi(y) b^\Lambda(y) T^\Lambda_a),
\quad\hbox{for}\quad
x\in [0, 2\pi],
\label{G16}\ee
where $Q_{F_\phi}$ is determined as described above.
By combining the preceding formulae,  one can verify that 
\be
\{ F_{\chi}, F_{\phi}\}=
\Omega_{Bloch}^{G,\kappa}(Y^{F_{\phi}},Y^{F_{\chi}})=
\int_0^{2\pi} \int_0^{2\pi} dx dy  \tr_{12}\left( \chi(x)\otimes \phi(y) 
\{ b^\Lambda(x) \stackrel{\otimes}{,} b^\Lambda(y)\}\right)
\label{locB}\ee
holds for any $\phi$, $\chi$ subject to (\ref{G13}) {\em provided that one has} 
\be 
\Big\{b^\Lambda(x)\stackrel{\otimes}{,} b^\Lambda(y)\Big\}
={1\over\kappa}\Big(b^\Lambda(x)\otimes
b^\Lambda(y)\Big)\Big({\cal R}(\omega) 
+{1\over 2}{\hat I} \,{\rm sign}\,(y-x)\Big)^\Lambda, 
\quad 0< x,y<2\pi
\label{xcharep}\ee 
with ${\cal R}(\omega)$ given by (\ref{Romega}).
Since the representation $\Lambda$ is arbitrary, 
this can be symbolically written in the form (\ref{xcha}).
It is clear that  
the local formula (\ref{xcha})
completely encodes the Poisson brackets 
since $Y^{F_{\phi}}$ can be recovered 
if the right hand side of  (\ref{locB}) is given.

Since the current $J=\kappa b' b^{-1}$ and the monodromy parameter 
$\omega$ are functions of $b$, their Poisson brackets can be derived from 
the exchange algebra (\ref{xcha}).
We can also determine  the hamiltonian vector 
fields of the functions $\omega_k := \tr(\omega H_k)$ and
$\F_\mu:=\int_0^{2\pi}dx{\tr}\Big(\mu(x)J(x)\Big)$,
where $\mu$ is a $2\pi$-periodic, smooth, $\G$-valued test function,
directly from (\ref{G1}), (\ref{G2}) as 
\be
Y^{\omega_k} (b(x))={1\over \kappa} b(x) H_k,
\qquad
Y^{\F_\mu}\big(b(x)\big)=\mu(x)b(x).
\ee
This implies that $J$ generates the KM Poisson 
brackets (\ref{JPB}),
and the current algebra is centralized by the functions of $\omega$.

It is worth remarking that the Jacobi identity 
of the Poisson bracket (\ref{xcha}) 
is equivalent to the following equation 
for the exchange  r-matrix: 
\be
[\cR_{12}(\omega), \cR_{23}(\omega)] +
 \sum_k H_1^k  {\pa \over \pa \omega^k} 
\cR_{23}(\omega) + \hbox{cycl.~perm.}  =-{1\over 4} 
f_{ab}^{\phantom{ab}c} T^a \otimes T^b \otimes T_c,
\label{CDYB}\ee
where  $[T_a, T_b]=f_{ab}^{\phantom{ab}c}T_c$ and 
$\tr(T_a T^b)=\delta_a^b$.
This is a dynamical generalization of the modified classical Yang-Baxter 
equation, and it has been verified in \cite{BDF} 
that ${\cal R}(\omega)$ in (\ref{Romega}) satisfies it.
This equation appears in other 
contexts as well \cite{Gervais,Feld,Avan} and was recently studied  
e.g.~in \cite{EV,Liu,Lu}.

The classical exchange
algebra (\ref{xcha}) is also valid
for a compact simple Lie group, $K$, obtained as a real form 
of a complex simple Lie group, $G$ (like $K=SU(n) \subset SL(n,{\bf C})=G$). 
To see this, let ${\cal K}\subset {\cal G}$ be a compact real form 
of a complex simple Lie algebra ${\cal G}$.
One can realize ${\cal K}$ as 
\be
{\cal K}={\rm{span}}_{\bf R}\{ i H_{\alpha_k}, 
F^+_\alpha, F^-_\alpha\,\vert\, \alpha_k\in \Delta,\,\,\, 
\alpha\in \Phi^+\,\} 
\ee
with
\be
F_\alpha^-={\vert\alpha\vert \over 2} (E_\alpha - E_{-\alpha}),
\quad
F^+_\alpha=i {\vert\alpha\vert\over 2}(E_\alpha + E_{-\alpha})  
\ee
where $H_{\alpha_k}$ ($\alpha_k\in \Delta$),  $E_{\pm \alpha}$ 
($\alpha\in \Phi^+$)
form a Chevalley basis of $\G$ 
corresponding to the set of simple roots, $\Delta$, 
and positive roots, $\Phi^+$ (see e.g.~\cite{Lie}).
If $\cal A$ is now a regular domain in the Cartan subalgebra of ${\cal K}$, 
we have  
\be
\A\ni \omega=i \tau = i \sum_k \tau^k H_{\alpha_k} 
\quad\hbox{with}\quad
\tau^k \in {\bf R}.
\ee
Using also that $\cot(y)= i \coth\left(i y\right)$, 
${\cal R}(\omega)$ in (\ref{Romega}) can be rewritten as 
\be
{\cal R}(i\tau)=  \sum_{\alpha \in \Phi^+} 
{\vert \alpha\vert^2\over 4} \coth(i{1\over 2} 
\alpha(\tau))  E_\alpha\wedge E_{-\alpha}
={1\over 2} \sum_{\alpha\in \Phi^+} \cot({1\over 2}\alpha(\tau))
 F_\alpha^+ \wedge F^-_\alpha.
\label{Rtau}\ee
This means that if $\omega=i\tau$, 
then the r-matrix on the right hand side of (\ref{xcha}) lies in 
${\cal K}\otimes {\cal K}$.
Therefore (\ref{xcha}) with (\ref{Romega})
can be consistently applied to  $K$-valued Bloch waves.
This can be shown to be the correct result 
in the compact case by deriving  
the exchange algebra similarly as above starting with 
 $\Omega^{K,\kappa}_{Bloch}$.
However, the free field realizations 
discussed subsequently are 
valid only in the case of a complex Lie algebra 
and its normal real form.

\section{Generalized Wakimoto realizations}
\setcounter{equation}{0}

At the classical level the generalized Wakimoto realizations  
of the current algebra are given by certain Poisson maps
\be 
W: (\widehat{\G}_0)^*_\kappa \times T^*\widetilde{G}_+ 
 \longrightarrow  (\widehat{\G})^*_\kappa.
\label{W}\ee
Here  $(\widehat{\G})^*_\kappa$ denotes the space  of $\G$-valued currents,
\be  
(\widehat{\G})^*_\kappa =\{ J\in C^\infty(\R, \G)\,\vert\, J(x+2\pi)= J(x)\,\},
\ee
equipped  with 
the Poisson bracket in (\ref{JPB}).
The other notations are explained below.
By elaborating the idea outlined in \cite{praga},
we will then prove that $W$ can be lifted to a Poisson 
(in fact, symplectic) map  
\be
\widehat W: \M^{G_0}_{Bloch} \times T^*\widetilde{G}_+ 
\longrightarrow  \M^{G}_{Bloch},
\label{hatW}\ee
which gives a realization of $G$-valued Bloch waves 
in terms of $G_0$-valued Bloch waves and free fields.

Let $\G=\sum_{n\in {\bf Z}} \G_n$ be an integral gradation of $\G$.
Define $\G_+:=\oplus_{n>0} \G_n$, $\G_-:=\oplus_{n<0} \G_n$.
Denote by $G_{0,\pm}\subset G$ the connected subgroups corresponding 
to $\G_{0,\pm}$.
One can associate a map $W$ (\ref{W}) 
with the parabolic subalgebra $({\cal G}_++\G_0)\subset \G$ as follows.
One of the constituents is the space 
\be
(\widehat{\G}_0)^*_\kappa =
\{ i_0\in C^\infty({\bf R}, \G_0)\,\vert\, i_0(x+2\pi)=i_0(x)\,\}
\ee
with the Poisson bracket  
\be
\{ \tr(t_k i_0)(x), \tr(t_l i_0)(y)\}= \tr([t_k, t_l] i_0)(x)\delta(x-y) 
+ \kappa\tr(t_k t_l) \delta'(x-y),
\label{i0PB}\ee
where $t_k$ denotes a basis of $\G_0$ and $\tr$ is the 
restriction of the scalar product of $\G$ to $\G_0$. 
To describe the other factor in (\ref{W}),
consider the manifolds $\widetilde{G}_+$ and $\widetilde{\G}_-$
whose elements are smooth, $2\pi$-periodic functions on $\R$ with values
in $G_+$ and $\G_-$, respectively.
By means of left translations, identify the 
cotangent bundle
of $\widetilde{G}_+$ as 
\be 
T^* \widetilde{G}_+ = \widetilde{G}_+ \times \widetilde{\G}_- =
\{ (\eta_+, i_-)\,\vert\, \eta_+\in \widetilde{G}_+,\,\,\, 
i_-\in \widetilde{\G}_-\,\}.
\ee
The  canonical symplectic form on $T^* \widetilde{G}_+$ is given by
\be
\Omega_{T^* \widetilde{G}_+}=
- d \int_0^{2\pi}dx \tr ( i_- \eta_+^{-1} d \eta_+),
\ee
and the corresponding Poisson brackets are encoded by  
\begin{eqnarray}
&&\{\,\tr(V^\alpha\, i_-)(x) \,, \, \tr(V^\beta\, i_-)(y) \,\}
     = \tr([V^\alpha,V^\beta]\, i_-)(x) \,\delta(x-y) ,
\nonumber \\
&& \{\,\tr (V^\alpha\, i_-)(x) \,, \, \eta_+(y) \,\}
     =  \eta_+(x)V^\alpha \,\delta(x-y) ,
\quad
\{ \eta_+(x)\stackrel{\otimes}{,} \eta_+(y)\}=0,
\label{iPB}\end{eqnarray}
where $V^\alpha$ is a basis of $\G_+$.
The map $W$ is defined by the formula:
\be
W: (\widehat{\G}_0)^*_\kappa \times T^* \widetilde{G}_+
\ni (i_0, \eta_+, i_- ) \mapsto J=\eta_+(i_0 -i_- ) \eta_+^{-1} + \kappa 
\eta_+' \eta_+^{-1}\in (\widehat{\G})^*_\kappa.
\label{classwak}\ee 
One can verify \cite{dBF} that this is a Poisson map, 
i.e., the Poisson bracket  of $J$ in  (\ref{JPB}) follows from the  Poisson 
brackets of the constituents $(i_0, \eta_+, i_-)$.

Our purpose now is to complete the construction of the following 
commutative diagram:
\be\begin{array}{ccc}
\M^{G_0}_{Bloch} \times T^*\widetilde{G}_+ &
\ \stackrel{{\widehat{W} }}{\Longrightarrow}\  
& \M^{G}_{Bloch} \\ {} & {} &\\
{\scriptstyle{{\cal D}_0 \times \rm{id}}}\,\, 
\downarrow {\phantom{\scriptstyle{{\cal D}_0 \times \rm{id}}}}
 & {} &  {\phantom{\scriptstyle{{\cal D}}}} \downarrow 
\,\, {\scriptstyle{{\cal D}}} \\ {} & {} & \\
(\widehat{\G}_0)^*_\kappa \times T^*\widetilde{G}_+ &
\ \stackrel{{W}} \longrightarrow\  & (\widehat{\G})^*_\kappa \\
\end{array}\label{diagram}\ee
The map  $\cal D$ operates according to 
${\cal D}: b \longmapsto J=\kappa b' b^{-1}$.
$\M^{G_0}_{Bloch}$ is the space of $G_0$-valued
Bloch waves with regular, diagonal monodromy,
\be
\M^{G_0}_{Bloch}=\{ \eta_0 \in C^\infty(\R, G_0)\,\vert\,
\eta_0(x+2\pi)= \eta_0(x) e^{\omega},
\quad 
\omega\in \A \subset \H\,\} 
\label{G0Bloch}\ee
 with the symplectic form
\be
\Omega^{G_0,\kappa}_{Bloch}(\eta_0)= 
- {\kappa\over 2 }\int_0^{2\pi}dx\,
 {\tr}\left(\eta_0^{-1}d\eta_0 \right) \wedge
\left(\eta_0^{-1}d\eta_0 \right)'
- {\kappa\over 2 }  {\tr}\left((\eta_0^{-1} d\eta_0)(0)\wedge d \omega\right).
\ee
Since the same domain $\cal A$ is used in (\ref{Bloch}) and (\ref{G0Bloch}),
$\M^{G_0}_{Bloch}$ is a symplectic submanifold of $\M^{G}_{Bloch}$.
The map $\D_0$ sends $\eta_0$ to $i_0=\kappa \eta'_0 \eta_0^{-1}\in 
(\widehat{\G}_0)^*_\kappa$.
We have seen that 
the simple arrows in (\ref{diagram}) are Poisson maps.
The formula of the missing map  
\be
\widehat{W}: \M^{G_0}_{Bloch} \times T^*\widetilde{G}_+ \ni
(\eta_0, \eta_+, i_-) \longmapsto b\in \M^{G}_{Bloch}
\ee
can be found from the equation 
$\D \circ \widehat{W}= W \circ (\D_0 \times {\rm id})$, 
which requires that
\be
\kappa b'b^{-1}= \eta_+(\kappa \eta_0'\eta_0^{-1} -i_- ) 
\eta_+^{-1} + \kappa \eta_+' \eta_+^{-1}.
\label{wakidiff}\ee
A solution for $b$ exists that admits a generalized Gauss decomposition.
In fact,  
\be
b(x)=b_+(x) b_0(x) b_-(x)
\quad\hbox{with}\quad  b_{\pm, 0}(x)\in G_{\pm, 0}
\ee
is a solution of (\ref{wakidiff}) if
\be
b_+=\eta_+, 
\quad
b_0=\eta_0
\quad
\hbox{and}\quad
\kappa b_-' b_-^{-1}= -  \eta_0^{-1} i_- \eta_0.
\label{bmeq}\ee
The general solution of the differential equation for $b_-$ 
can be written in terms of the particular solution 
$b_-^P$, defined by  
$b_-^P(0)={\bf 1}$ as $b_-(x)= b_-^P(x)S$ with an arbitrary  
$S\in G_-$. (Note that $b_-^P(x)$ is nothing but the
path ordered exponential integral of $-  \eta_0^{-1} i_-
\eta_0/\kappa$ over $[0,x]$.) 
The constant $S=b_-(0)$ has to be determined from the condition 
that $b$ should have diagonal monodromy.
One finds that $b$ has diagonal monodromy,
indeed it satisfies $b(x+2\pi)= b(x) e^{ \omega}$,
if and only if 
\be
e^{-\omega} S e^{ \omega } = b_-^P(2\pi) S.
\label{Seq}\ee
Inspecting this equation grade by grade 
using  a parametrization $S=e^s$, $s\in \G_-$ and 
the fact that $\omega\in \A$, 
it is not difficult 
to see that it has 
a unique solution for $S$ as a function of $\omega$ and $b_-^P(2\pi)$. 
Determining $S$ in the above manner, we now define the  map
\be
{\widehat W}: 
\M^{G_0}_{Bloch} \times T^*\widetilde{G}_+ \ni
(\eta_0, \eta_+, i_-) \longmapsto b=\eta_+ \eta_0 b_-^P S 
\in \M^{G}_{Bloch}
\label{wakiext}\ee
that makes the diagram in (\ref{diagram}) commutative.

The main result of this paper is the following statement:
{\em The map $\widehat{W}$ defined in (\ref{wakiext})
is symplectic, that is, }
\be
(\widehat{W}^* \Omega^{G, \kappa}_{Bloch})(\eta_0, \eta_+, i_-)=
\Omega^{G,\kappa}_{Bloch}(b=\eta_+ \eta_0 b_-^P S)=
\Omega^{G_0,\kappa}_{Bloch}(\eta_0)
+\Omega_{T^*\widetilde{G}_+}(\eta_+,i_-).
\label{CLAIM}\ee

To prove this,
let us restrict the symplectic form $\Omega_{Bloch}^{G,\kappa}$
in (\ref{Blochform}) to the domain of $\M^G_{Bloch}$ whose 
elements are decomposable in the form
\be
b=b_+ b_0 b_- 
\qquad\hbox{with} \qquad b_{0,\pm} \in C^\infty(\R, G_{0,\pm}),
\label{+*}\ee
\be
b_+(x+2\pi)=b_+(x),
\qquad
b_0(x+2\pi)=b_0(x)e^\omega,
\qquad
b_-(x+2\pi)=e^{-\omega} b_-(x) e^\omega.
\label{+**}\ee
A straightforward calculation that uses partial integration 
and standard properties of the trace yields that in this domain 
\bea
&&\Omega^{G,\kappa}_{Bloch}(b)= 
- {\kappa\over 2 }\int_0^{2\pi}dx\,
 {\tr}\left(b_0^{-1}d b_0  \wedge (b_0^{-1}db_0 )'
+2 (b_0 b_-' b_-^{-1} b_0^{-1})
(b_+^{-1} db_+ \wedge b_+^{-1} db_+)\right) 
\nonumber\\
&&\quad
+ \kappa\int_0^{2\pi}dx\, \tr\left( 
b_0 b_- (b_-^{-1} db_-)' b_-^{-1} b_0^{-1} \wedge b_+^{-1} db_+ 
+[ db_0 b_0^{-1} , b_0 b_-' b_-^{-1} b_0^{-1} ]\wedge b_+^{-1} db_+\right)
\nonumber \\
\nonumber
&&\quad
- {\kappa\over 2 }  {\tr}\left((b_0^{-1} d b_0)(0)\wedge d \omega\right)
-{\kappa\over 2} \tr\left( 
\left(b_-^{-1} b_0^{-1}  (b_+^{-1} db_+) b_0 b_-\right)(0)
\wedge d\omega\right)
\nonumber\\
&&\quad
-{\kappa\over 2}\left[ \tr\left( b_-^{-1} db_- \wedge b_-^{-1} b_0^{-1} 
(b_+^{-1} db_+) b_0 b_-\right)\right]_0^{2\pi}. 
\eea 
The last two terms cancel, and combining also the other terms we get
\bea
&&\Omega^{G,\kappa}_{Bloch}(b)= 
- {\kappa\over 2 }\int_0^{2\pi}dx\,
{\tr}\left(b_0^{-1}d b_0 \right) \wedge
\left(b_0^{-1}db_0 \right)'
- {\kappa\over 2 }  {\tr}\left((b_0^{-1} d b_0)(0)\wedge d \omega\right)
\nonumber\\
&&\qquad\qquad 
-d \int_0^{2\pi}dx\, \tr\left( 
(-\kappa b_0 b_-' b_-^{-1} b_0^{-1})  b_+^{-1} db_+ \right).
\nonumber 
\eea
The definition of 
$\widehat{W}$ says that the image   of 
$(\eta_0, \eta_+, i_-) 
\in\M^{G_0}_{Bloch} \times T^*\widetilde{G}_+ $ 
is  $b \in\M^{G}_{Bloch}$ that has the form 
(\ref{+*}) with $b_0=\eta_0$, $b_+=\eta_+$ and $b_-$ specified by  
the monodromy condition in (\ref{+**}) together with 
$-\kappa \eta_0 b_-' b_-^{-1} \eta_0^{-1} = i_-$.
Taking this into account, the last equation immediately implies that 
\be
(\widehat{W}^* \Omega^{G, \kappa}_{Bloch})(\eta_0, \eta_+, i_-)=
\Omega^{G_0,\kappa}_{Bloch}(\eta_0) 
- d \int_0^{2\pi}dx\, \tr ( i_- \eta_+^{-1} d \eta_+),
\ee
as required by (\ref{CLAIM}).

Since a symplectic map is always  a Poisson map as well,
$\widehat{W}$ provides us with a realization of the monodromy 
dependent exchange algebra (\ref{xcha}) of the $G$-valued Bloch waves 
in terms of the analogous exchange algebra of the $G_0$-valued 
Bloch waves and 
the Poisson algebra of $T^*\widetilde{G}_+$ given by (\ref{iPB}).
Of course,  $T^*\widetilde{G}_+$ can also be parametrized
by canonical free fields. 
For this, consider some global coordinates $q^\alpha$ on $G_+$ 
and define the matrix   
${\cal N}_{\alpha\beta}(q)=\tr\left(V_\beta \eta_+^{-1}
{\pa  \eta_+  \over \pa q^\alpha}
\right)$,
where $V_\beta$ is a basis of $\G_-$ dual to the 
basis $V^\alpha$ of $\G_+$.
By introducing $2\pi$-periodic canonical free 
fields $q^\alpha(x)$, $p_\beta(y)$,
\be
\{ q^\alpha(x), p_\beta(y)\}=\delta^\alpha_\beta,
\qquad
\{ q^\alpha(x), q^\beta(y)\}=\{ p_\alpha(x), p_\beta(y)\}=0,
\ee
the Poisson brackets in (\ref{iPB}) are realized by  $\eta_+(x)=
\eta_+(q(x))$ and 
\be
i_-(x)=- \sum_{\alpha\beta} 
({\cal N}^{-1})^{\alpha\beta}(q) p_\beta V_\alpha,
\ee
whereby $\Omega_{T^* \widetilde{G}_+}=\int_0^{2\pi}dx\, 
(dp_\alpha\wedge d q^\alpha)$.
The map $\widehat{W}$ gives a true free field
realization in the principal
case, for which $\G_0=\H$ is Abelian.
In this case  $\eta_0$ is the exponential
of a $\H$-valued free scalar field, $\Psi(x)$, i.e., 
$\eta_0(x)=e^{\Psi(x)}$ with $\Psi(x+2\pi) =\Psi(x) + \omega$ and
\be
\{ \tr(H_k \Psi)(x), \tr(H_l\Psi)(y)\}  =
{1\over 2\kappa} \tr(H_k H_l) \,{\rm sign}\,(y-x) 
\quad\hbox{for}\quad 0< x, y<2\pi.
\ee
For $G=SL(2)$ the Wakimoto realization 
of Bloch waves  was already described in 
\cite{AS,FG} and some other special cases can be found in 
\cite{GMOMS}.
These results are in agreement with our general construction
of the map $\widehat W$ in (\ref{wakiext}).

Finally, we illustrate the formula in (\ref{wakiext}) 
by a series of simple examples  
for the group $G=SL(n)$ (either real or complex).
The parabolic subalgebras of $sl(n)$ are associated with the partitions 
of $n$, and we consider the 2-block cases
(the map $W$ in (\ref{W}) is described in these cases in \cite{dBF}). 
That is to say, we let the underlying integral gradation of $\G=sl(n)$
be defined by the eigenvalues of $\ad Q$ with a diagonal matrix
\be
Q={1\over n}{\rm diag}\left( n_2 {\bf 1}_{n_1}, -n_1 
{\bf 1}_{n_2}\right),
\qquad 
n=n_1+n_2.
\ee
In this case $\G_\pm$ are Abelian subalgebras, which leads to
a simplification of the formulae.
In particular, we can introduce the convenient parametrizations
\be
\eta_+ =\left[ \matrix{ {\bf 1}_{n_1} & q \cr 0 & {\bf 1}_{n_2} \cr}\right],
\qquad
i_- =-\left[ \matrix{ {\bf 0}_{n_1} & 0 \cr p & {\bf 0}_{n_2} \cr}\right],
\ee
where $q(x)$ and $p(x)$ are $n_1\times n_2$ and $n_2\times n_1$ matrices,
whose entries satisfy
\be
\{ q_{ab}(x), p_{cd}(y)\}= \delta_{ad} \delta_{bc} \delta(x-y).
\ee
In terms of the parametrizations  
\be
\eta_0 = 
\left[ \matrix{ {\eta}_{u} & 0 \cr 0 & 
\eta_{d} \cr}\right],
\qquad
b^P_-= 
\left[ \matrix{ {\bf 1}_{n_1} & 0 \cr B & {\bf 1}_{n_2} \cr}\right],
\ee
the solution of the differential equation in (\ref{bmeq})
is then found to be  
\be
B(x)={1\over \kappa} \int_0^x dy\, \left(\eta_d^{-1}
p \eta_u\right)(y).
\ee
Furthermore, if we now denote 
\be
S= \left[ \matrix{ 
{\bf 1}_{n_1} & 0 \cr \sigma & {\bf 1}_{n_2} \cr}\right],
\qquad
\omega =
{\mathrm diag}\left(\omega_1,\omega_2, \ldots, \omega_n\right),
\ee
then eq.~(\ref{Seq}) is explicitly solved as
\be
\sigma_{ab}= { B_{ab}(2\pi) \over 
\exp(\omega_b - \omega_{a + n_1})  -1},
\qquad
1 \leq a \leq n_2,
\,\,\, 
1\leq b \leq n_1.   
\ee
Combining these equations, 
(\ref{wakiext}) yields  a realization of the $SL(n)$-valued
Bloch-wave $b$ in terms of the fields $q$, $p$ and $\eta_0$.
By subsequently using a similar Wakimoto realization 
for $\eta_{0}$, and so on, one can iteratively
build up a complete free field realization of $b$.
Incidentally, $\eta_0$ can  be written in the alternative form
\be
\eta_0 = e^{Q\varphi }
\left[ \matrix{ \tilde{\eta}_{u} & 0 \cr 0 & 
\tilde\eta_{d} \cr}\right],
\ee
where $e^{Q\varphi(x)}$, $\tilde \eta_u(x)$ and $\tilde \eta_d(x)$
are  $U(1)$, $SL(n_1)$ and $SL(n_2)$-valued 
independent Bloch-waves. 

\medskip

The Wakimoto realizations of the affine KM algebras in 
generalized Fock spaces, at the level 
of vertex algebras  as opposed to the above
Poisson algebras, have many applications in conformal field theory
\cite{DiF,FF,BMP}. 
An explicit formula for such realizations
of the current $J$  was derived in \cite{dBF} by 
quantizing the expression (\ref{classwak}).
It would be very interesting to also quantize (\ref{wakiext}).
The exchange algebra of the resulting 
vertex operators should contain the quantized version of the 
r-matrix ${\cal R}(\omega)$ which has been constructed recently 
for all Lie algebras  in a universal manner \cite{Arnaudon}.
Another open problem is to find an analogue 
of the rather simple construction of the Wakimoto realizations 
presented here and in \cite{dBF} for the case of $q$-deformed
affine KM algebras.
We hope to return to these questions in the future.

\bigskip
\bigskip
\noindent
{\bf Acknowledgements.}
This investigation was supported in part by the Hungarian National
Science Fund (OTKA) under T019917, T030099, T025120 
and by the Ministry
of Education under FKFP 0178/1999 and FKFP 0596/1999.


\begin{thebibliography}{99}

\parskip=2.0pt

\bibitem{Witt}
E. Witten, Commun. Math. Phys.  92 (1984) 455.
\bibitem{DiF}
P. Di Francesco, P. Mathieu and D. S\'en\'echal, 
Conformal Field Theory (Springer, 1996).
\bibitem{EFV} 
P. Etingof, I. Frenkel and A. Kirillov Jr., 
Lectures on Representation Theory and Knizhnik-Zamolodchikov Equations
(AMS, 1998).
\bibitem{Bab}
O. Babelon, Phys. Lett. B 215 (1988) 523.
\bibitem{Blok}
B. Blok, Phys. Lett.  B 233 (1989) 359.
\bibitem{Fad}
L. Faddeev, Commun. Math. Phys. 132 (1990) 131.
\bibitem{AS}
A. Alekseev and S. Shatashvili, Commun. Math. Phys. 133 (1990) 353.
\bibitem{IY}
T. Itoh and Y. Yamada, Prog. Theor. Phys. 85 (1991) 751.
\bibitem{FL}
V.A. Fateev and S.L. Lukyanov, Int. J. Mod. Phys. A 7 (1992) 853.
\bibitem{AT}
A. Alekseev and I.T. Todorov, Nucl. Phys. B 421 (1994) 413.
\bibitem{BDF}
J. Balog, L. D\c{a}browski and L. Feh\'{e}r, 
Phys. Lett. B 244 (1990) 227. 
\bibitem{Chu5}
M. Chu, P. Goddard, I. Halliday, D. Olive and A. Schwimmer,
    Phys. Lett. B 266 (1991) 71.
\bibitem{BBT}
O. Babelon, F. Toppan and L. Bonora, Commun. Math. Phys. 140 (1991) 93.
\bibitem{Gaw}
K. Gaw\c{e}dzki, Commun. Math. Phys. 139 (1991) 201.
\bibitem{FG}
F. Falceto and K. Gaw\c{e}dzki, 
J. Geom. Phys. 11 (1993) 251.
\bibitem{BFP}
J. Balog, L. Feh\'er and L. Palla, 
Phys. Lett. B 463 (1999) 83 (to appear), {\tt hep-th/9907050};  
Chiral extensions of the WZNW phase space, Poisson-Lie symmetries
and groupoids, {\tt hep-th/9910046}.
\bibitem{qgroup}
C. G\'omez, M. Ruiz-Altaba and G. Sierra,
Quantum Groups in Two-Dimensional Physics 
(Cambridge University Press, 1996), and references therein. 
\bibitem{Chu2}
M. Chu and P. Goddard, Phys. Lett. B 337 (1994) 285; 
 Nucl. Phys. B 445 (1995) 145.
\bibitem{FHT}
P. Furlan, L.K. Hadjiivanov and  I. T. Todorov,
Nucl.Phys. B 474 (1996) 497;  Int. J. Mod. Phys. A 12 (1997) 23.
\bibitem{CL}
L. Caneschi and M. Lysiansky, Nucl. Phys. B 505 (1997) 701. 
\bibitem{Wa}
M.~Wakimoto, 
Commun.\ Math.\ Phys.\ {\bf 104} (1986)  605.
\bibitem{FF}
B.L.~Feigin and  E.V.~Frenkel,
Russ.\ Math.\ Surv.\ {\bf 43} (1989)  221; 
Commun.\ Math.\ Phys.\ {\bf 128} (1990) 161;\\
E.~Frenkel, 
Free field realizations in representation theory and conformal field theory,
{\tt hep-th/9408109}.
\bibitem{BMP}
P.~Bouwknegt, J.~McCarthy and  K.~Pilch,
Commun.\ Math.\ Phys.\ {\bf 131} (1990) 125;
Prog.\ Theor.\ Phys.\ Suppl.\ {\bf 102} (1990)  67.
\bibitem{GMOMS}
A.~Gerasimov, A.~Morozov, M.~Olshanetsky, A.~Marshakov and  S.~Shatashvili,
Int.~J.\ Mod.\ Phys.\ {\bf A5} (1990) 2495.
\bibitem{dBF}
J.~de Boer and L.\ Feh\'er, 
Mod. Phys. Lett. A {\bf 11} (1996) 1999;
Commun.\ Math.\ Phys. {\bf 189} (1997) 759.
\bibitem{praga}
L. Feh\'er, in: Proc. of the 7th Colloquium on Quantum Groups and
Integrable Systems, Czech. J. Phys. {\bf 48} (1998) 1325.
\bibitem{Gervais}
J.-L. Gervais and A. Neveu,  Nucl. Phys. B 238 (1984) 125;\\
E. Cremmer and J.-L. Gervais, Commun. Math. Phys. 134 (1990) 619.
\bibitem{Feld}
G. Felder and C. Wieczerkowski, Commun. Math. Phys. 176 (1996) 133;\\
G. Felder, pp.~1247-1255 in: Proc. Int. Congr. Math. Z\"urich,   1994
{\tt (hep-th/9407154)}. 
\bibitem{Avan}
J. Avan, O. Babelon and E. Billey, Commun. Math. Phys. 178 (1996) 281.
\bibitem{EV} P. Etingof and A. Varchenko, 
Commun. Math. Phys. 192 (1998) 77.
\bibitem{Liu} Z. J. Liu and P. Xu, 
Dirac structures and dynamical r-matrices, {\tt math.DG/9903119}.
\bibitem{Lu}
J.H. Lu, Classical dynamical r-matrices and homogeneous Poisson 
structures on $G/H$ and $K/T$, {\tt math/9909004}.
\bibitem{Lie}
V.V.~Gorbatsevich, A.L.~Onishchik and E.B.~Vinberg, 
Structure of Lie Groups and Lie Algebras, 
Encyclopaedia of Mathematical Sciences, Vol.~41
(Springer,  1994).  
\bibitem{Arnaudon} 
D. Arnaudon, E. Buffenoir, E. Ragoucy and Ph. Roche,
Lett. Math. Phys. {\bf 44} (1998) 201.

\end{thebibliography}
\end{document}